\documentclass[11pt,twoside]{book}
\usepackage{konkolyproc2}
\usepackage{longtable}
\usepackage{amsmath,amssymb}
\usepackage{graphicx}
\usepackage{lscape}
\usepackage{index}
\usepackage{natbib}
\usepackage{bigdelim}
\usepackage{multirow}
\makeindex

\begin{document}

%%%%%%%%%%%%%%%%%%%%%%%%%%%%%%%%%%%%%%%%%%%%%%%%%%%%%%%%%%%%%%%%%%%%%%%%%%
\pagestyle{myheadings}
\setcounter{equation}{0}\setcounter{figure}{0}\setcounter{footnote}{0}\setcounter{section}{0}\setcounter{table}{0}\setcounter{page}{1}
\markboth{Szabados, Szab\'o \& Kinemuchi}{RRL2015 Conf. Papers}

\title{The magnificent past of RR Lyrae variables}
\author{Ennio Poretti$^{1,2,3,4}$, Jean-Francois Le Borgne$^{2,3,4}$, Alain Klotz$^{2,3,4}$, Maurice Audejean$^5$ \&  Kenji Hirosawa$^6$}
\affil{
$^1$INAF-Osservatorio Astronomico di Brera, Merate (LC), Italy\\
$^{2}$Universit\'e de Toulouse, UPS-OMP, IRAP, Toulouse, France\\
$^{3}$CNRS, IRAP,  Toulouse, France\\
$^{4}$GEOS, Groupe Europ\'een d'Observations Stellaires, France\\
$^{5}$Observatoire de Chinon, Astronomie en Chinonais, Chinon, France\\
$^{6}$Variable Star Observers League(VSOLJ), Matsushiro,  Japan
}
%%%%%%%%%%%%%%%%%%%%%%%%%%%%%%%%%%%%%%%%%%%%%%%%%%%%%%%%%%%%%%%%%%%%%%%%%%%

\begin{abstract}
We briefly review the results obtained by using the times of maximum brigthness of RR Lyr variables. 
They cover more than 120 years and have been used to study the evolutionary changes of the pulsational
periods, the different shapes of the Blazhko effect among galactic RRab stars, and the interplay between
pulsational and Blazhko periods in the eponym of the class, RR Lyr.
The data are stored in the GEOS database, continuously fed by contributions from professional and 
amateur astronomers.
\end{abstract}

\section{Introduction}
The history of the observations of RR Lyr variables started  in the XIXth century,
more than 120 years ago.
The variability of RR~Lyr itself  was discovered on the photographic plates of the
{\it Henry Draper Memorial}  by Mrs.~Williamina P. Fleming \citep{pickering}. The first measurement
comes  back to July 20, 1899, the first time of maximum brightness ($T_{\rm max}$)
to September 23, 1899. The list of $T_{\rm max}$ recorded from 1906 July 15 to August 25 
on RW Dra clearly shows variations of amplitude and phase \citep{blazhko}. \citet{shapley}
noted the same variations in the data of RR Lyr itself: this phenomenon was later named 
{\it Blazkho effect} in the astronomical literature.

The very long time baseline of available data combined with the short period of RR Lyrae 
variables offer an unique opportunity to look at their past as a treasure of valuable
information. At this purpose, 
the amateur/professional association Groupe Europ\'een d'Observations Stellaires (GEOS)
has built a database aimed to gather all the published $T_{\rm max}$ values \citep{pervar}. It 
also promotes the regular monitoring of RR Lyrae variables by means of own-made instruments
and observing time at robotic telescopes.

\section{Evolutionary changes}
\begin{figure}[!t]
\includegraphics[width=0.95\textwidth]{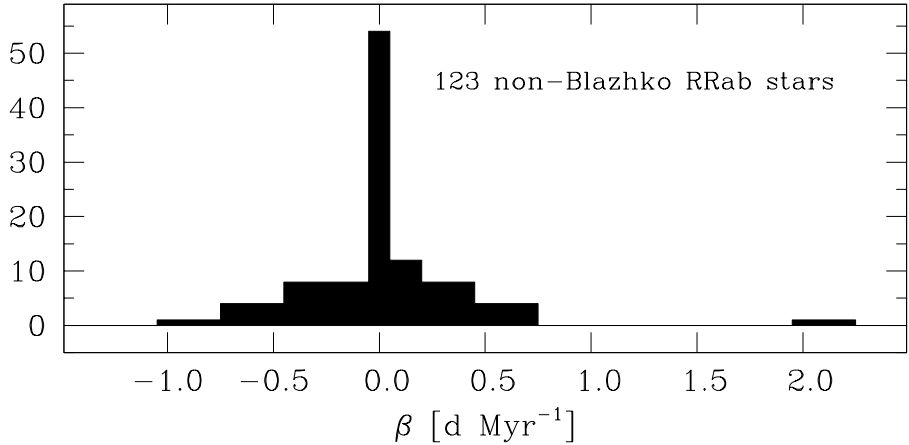}
\caption{Distribution of the rates of period changes. The extreme positive value is that of SV Eri}
\label{isto} 
\end{figure}

The long time baseline allowed us the opportunity to study period changes due
to stellar evolution \citep{pervar}. Most of the 123 scrutinized RRab stars does not
show any significant period variation. This reflects the fact that the rapid evolution is confined in short 
evolutionary phases. Notwithstanding this, we could put in evidence period increases in 27 stars
and decreases in 21 ones (Fig.~\ref{isto}).  The median values of the rates are +0.14~d~M~yr$^{-1}$ and 
--0.20~d~M~yr$^{-1}$, respectively. The order of magnitude of these rates is that expected from evolutionary
models and being common to a large number of stars it strongly supports the evolutionary origin.

We used stars not showing the Blazhko effect to study the evolutionary changes. Or, at least, not showing an
appreciable effect for the ground-based observations. Indeed, nowadays space-based time series are detecting very small
Blazhko effects, as in the case of V350 Lyr \citep{v350lyr}. 
Even if present in our sample, these small effects are not able to mask the large amplitude 
evolutionary  changes. 
Several case of erratic changes were observed, both increasing and decreasing the period. They should be 
the result of particular temporary instabilities rather than be considered as representative of a particular
evolutionary stage.
It was also noticed a slight excess of RR Lyr stars showing period decreases (i.e., stars on blueward evolutionary 
tracks). However, further studies on a larger sample suggested that the excess was mainly due to the limited
statistics \citep{geoscc}.

\section{The different shapes of the Blazhko effect}
We used the GEOS database to study the Blazhko effect of galactic RRab stars, too \citep{blagal}. 
The closed curves representing the Blazhko effect are constructed by plotting the magnitudes at maximum ($V_{\rm max}$)
vs. the $O-C$ ({\it observed} minus {\it calculated}) $T_{\rm max}$ values. 
We could emphasize some clear observational facts: {\it i)} the same value of the Blazhko period is observed at different values of the
pulsation periods; {\it ii)} different values of the Blazhko periods are observed at the same value of the pulsation
period; {\it iii)} the closed curves describing the Blazhko cycles have different shapes.  
These curves have often the shape of a potato, but other exotic contours are observed; 
{\it iv)} both clockwise and counterclockwise  are possible for closed curves with similar shapes.  

\begin{figure}[!t]
\includegraphics[width=0.88\textwidth]{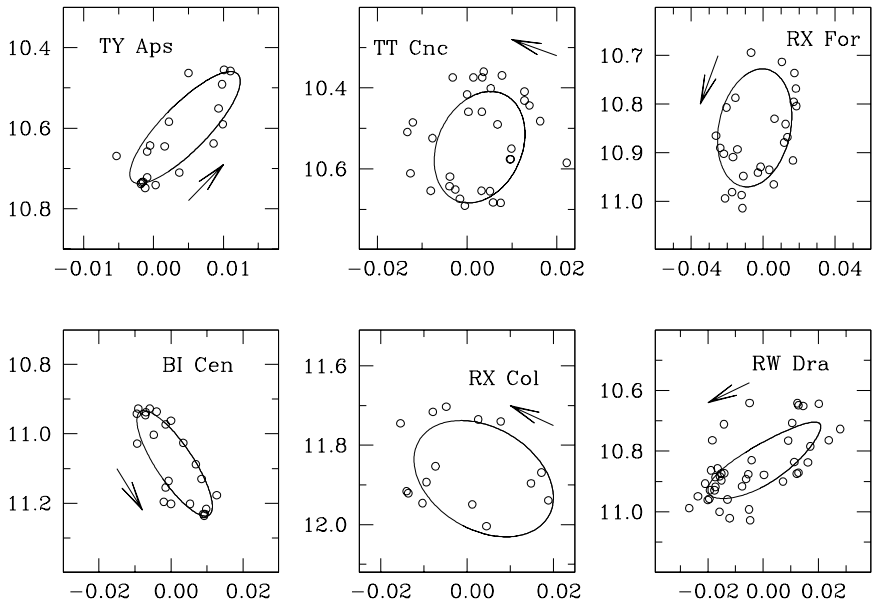}
\caption{The full rotation of the axis in stars running counterclockwise the closed curves in the O-C (abscissa, in days) vs.
$V_{\rm max}$ (ordinate) plots.}
\label{clock} 
\end{figure}

All these facts lead to a variegate group of behaviours. For instance,
the brightest maximum could correspond to the most negative phase shift in some stars and to the most positive one in 
others. We could observe a full rotation of the {\it Blazhko potato} in the sample of galactic RRab stars.
Figure~\ref{clock} shows the case of stars running counterclockwise. The regular survey of
RRab stars is undergoing using both amateurs' observatories and the robotic TAROT telescopes \citep{tarot,klotz,blagal}.

\section{The vanishing of the Blazhko effect of RR Lyr}
The analysis of the $T_{\rm max}$  epochs listed in the GEOS database allowed us to reconstruct
the changes in the pulsational period of RR~Lyr. We could establish the existence of two states
characterized by the pulsation over a  ``long" period  (longer than 0.56684~d) and over a ``short"
one (shorter than 0.56682~d). 
We also determined the Blazhko period in several time intervals since 1910
and we verified how it changed while the two states alternated. The variations of the pulsation  
and  Blazhko periods resulted  completely decoupled, since the Blazhko period  
had just one sudden decrease from 40.8~d to 39.0~d in 1975 \citep{itself}.
The current period is the shortest never measured, i.e., less than 0.5668~d. 
  
\begin{figure}[!t]
\includegraphics[width=1.0\textwidth]{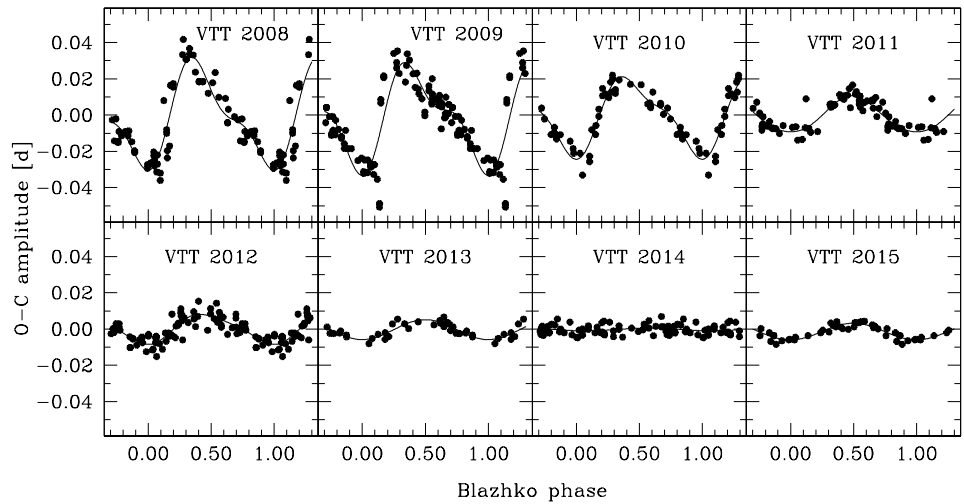}
\caption{O-C values of RR Lyr folded with the Blazhko period}
\label{vtt} 
\end{figure}

Some years ago, two of us (J.F.~Le Borgne and A.~Klotz) planned and realized small,
autonomous and transportable photometric instruments to make the ground-based survey of 
RR~Lyr as effective as possible. They assembled a commercial equatorial mount
(Sky-Watcher HEQ5 Pro Goto), an AUDINE CCD camera (512x768 kaf400 chip) and a
photographic 135-mm focal, f/2.8 lens with a field of view of 2$^\circ$\,x\,3$^\circ$.
They gave them the nickname VTTs for ``Very Tiny Telescopes". The regular recording of the 
$T_{\rm max}$'s of RR Lyr started in 2008. Data from VTTs and {\it Kepler} have been
extensively used to record the monotonic long-term decrease in the amplitude of the Blazhko effect.
As a matter of fact, the Blazhko effect was  hard to detect by looking at the maxima
collected in 2014 only. However, the new $T_{\rm max}$  collected with the VTTs in 2015 seem
to show a slight increase in the amplitude of the O-C values (Fig.~\ref{vtt}), a sign that the still unknown 
Blazhko mechanism is back at work.

\vskip 0.3truecm
{\bf Acknowledgments.} The projects exploiting the GEOS database have been developed during some visits of  E.~Poretti
at the {\it Institut de Recherche en Astrophysique et Plan\'etologie} in Toulouse. Supports from this Institute
are gratefully acknowledged.

\end{document}